\documentstyle[12pt]{article}
\topmargin = - 1 cm
\font\twelve=cmbx10 at 15pt
\font\ten=cmbx10 at 12pt

\renewcommand{\thefootnote}{\fnsymbol{footnote}}

\begin{document}

\begin{titlepage}

\begin{center}

{\ten Centre de Physique Th\'eorique\footnote{Unit\'e Propre de
Recherche 7061} - CNRS - Luminy, Case 907}

{\ten F-13288 Marseille Cedex 9 - France }

\vspace{4 cm}

{\twelve INVESTIGATION OF THE HIGH ENERGY BEHAVIOURS OF THE SCALAR PARTICLE
SCATTERING AMPLITUDE IN THE GRAVITATIONAL FIELD BY FUNCTIONAL
APPROACH\footnote{To submitted to Nuovo Cimento A.}}

\vspace{0.3 cm}
\setcounter{footnote}{0}
\renewcommand{\thefootnote}{\arabic{footnote}}

{\bf Nguyen Suan Han\footnote{Permanent address:
                 Department of Theoretical Physics, Hanoi State University,

                 \quad P. O. Box 600, BoHo, Hanoi, 10000, Vietnam}}

\vspace{1.5 cm}

{\bf Abstract}

\end{center}

Closed expressions for the Green function and amplitude of the
      scalar particle scattering in the external gravitational field
      $ g_{\mu\nu}(x)  $ are found in the form of functional integrals. It is
      shown that,  as compared with the scattering on the vector potential,
the
      tensor character of the gravitational field leads to a more rapid
      increase of the cross section with increasing energy.  Discrete energy
      levels of particles are obtained in the Newton potential.

\vspace{2 cm}


\vspace{2 cm}

\noindent November 1994

\noindent CPT-94/P.3084

\bigskip

\noindent anonymous ftp or gopher: cpt.univ-mrs.fr

\end{titlepage}

\section{Introduction}

	The method of solving equations for the one-particle Green function
      in the external field by  the functional integral has been
      given by Barbashov $ [ 1 ] $.  A closed  solution of the particle
      Green function is effective in studying  particle interactions in the
      infrared and high energy regions [2-4].
	In this note,  we present a generalization of this method to nonlinear
      interactions including the derivative coupling. We consider a scalar
field
      ${\Psi(x)} $ interacting with a gravitational field $ g_{\mu\nu} $.  The
      closed  solution of the scalar particle Green function in the
      external field $ g_{\mu\nu} $ has obtained. To contrast with the work
$[5]$
      where  the problem  has been considered an
      additional condition on the field $ g_{\mu\nu} $  /harmonic condition/
      is explicitly taken into account  in our method. The scalar particle
Green
function represented by
       the functional integral is used for  constructing the
      scattering amplitude. In the high energy region the eikonal
      representation for the scattering amplitude of the scalar particle in
      the tensor potential is obtained. The Newton potential is considered
      as an example.   It is shown that as compared with the
      scattering on vector potential,  the character of the gravitational field
      leads to a more rapid increase in the cross section with increasing
      energy. Discrete energy  levels of particles are obtained in the Newton
      potential.

\section{The scalar particle Green function in an external
       gravitational field}

	Let us consider the model of  interaction of a scalar field with
      a gravitational field $ g_{\mu\nu} $ where the interaction Lagrangian
      is of the form

$$ L(x) = \frac{\sqrt{-g}}{2} \left[ g^{\mu\nu}(x)\partial_{\mu}\Psi(x)
	 \partial_{\nu}\Psi(x) - m^2 \Psi(x)^2 \right]\ ,\eqno(1) $$
where

$$  g = detg_{\mu\nu} =\det\sqrt{-g}g^{\mu\nu}(x)\ .$$

      Variating the Lagrangian (1) leads to the following equation for
      field ${\Psi(x)}$:

$$\left[ -\tilde g^{\mu\nu}(x)\partial_{\mu}\partial_{\nu}-\sqrt{-g}m^2
- \partial_{\mu}\tilde g^{\mu\nu}(x)\partial_{\nu}\right]{\Psi(x)} = 0\ ,
\eqno(2)$$

$$ \tilde g^{\mu\nu}(x) = \sqrt{-g}g^{\mu\nu}(x)\ . $$

        Equation (2) is conveniently investigated in the harmonic
      coordinates defined by the condition
$$ \partial_{\mu}\tilde g^{\mu\nu}{(x)} = 0\ . \eqno(3) $$

	The harmonic gauge (3)  being the analogy of the Lorentz gauge in
      the electrodynamics has  led to eliminate  the nonphysical component
      of the tensor field. Taking Eq.(3) into account, Eq.(2) yields to:

$$\left[ \tilde g^{\mu\nu}(x) i\partial_{\mu}i\partial_{\nu}
- \sqrt{-g}m^2\right]{\Psi(x)} = 0 \ .$$

      For the scalar particle Green function in the gravitational field we
      have the following equation
$$\left[ \tilde g^{\mu\nu}(x)i\partial_{\mu}i\partial_{\nu} -
\sqrt{-g}m^2 \right]G(x, y|g^{\mu\nu}) = -\delta^{(4)}(x-y)\ .\eqno(4)$$
      Equation (4) can be written in an operator form, if one uses the
      representation of the inverse operator $ \left [ g^{\mu\nu}(x)i
      \partial_{\mu}i\partial_{\nu} -\sqrt{-g} m^2 \right ]^{-1} $
      proposed by Fock and Feynman $ [7,8] $ as an exponent form
$$ G(x, y|g^{\mu\nu}) =$$
$$ i\int_{0}^{\infty}{d\tau}{\exp\left( -im^2
\int_{0}^{\tau}\sqrt{-g(x, \xi)}{d\xi} +
i\int_{0}^{\tau}\tilde g^{\mu\nu}(x, \xi)i\partial_{\mu}(\xi)
i\partial_{\nu}(\xi){d\xi}\right)}\delta^{(4)}(x-y)\ . \eqno(5) $$

      In this notation an exponent, whose the coefficient has noncommuting
      quantities as ${\partial_{\mu}}(x)$,  ${\tilde g^{\mu\nu}} $ and
      ${ g(x)} $,  is considered as $ T_{\xi} $-exponent, where   ${\xi}$ plays
      the role of an ordering index.   The coefficient of the exponent in the
      Eq.(5) is quadratic in the differentiation operator ${
      \partial_{\mu} }$.   However,  the transition from $T_{\xi}$-exponent to
       an  ordinary operator expression /"disentangling" the operators by the
      terminology of Feynman/ can not be performed without the series
      expansion. But one can  lower the power of the operator
      ${{\partial}_{\mu}} $ in  Eq.(5) if one uses the following formal
      transformation $[1]$:

 $$\exp\left( i\int_{0}^{\tau} {d\xi}\tilde g^{\mu\nu}(x, \xi)
 i\partial_{\mu}(\xi)i\partial_{\nu}(\xi)\right) =  $$
$$  C_{\nu}\int\prod_{\scriptstyle{\eta}}{d^4{\nu}(\eta)}exp\left(
 i\int_{0}^{\tau}{[\tilde g^{\mu\nu}(x, \xi)]^{-1}}
 {\partial_{\nu}(\xi)}{\partial_{\nu}(\xi)} -
  2i\int_{0}^{\tau}{d\xi}{\nu}^{\mu}(\xi){\partial_{\mu}(\xi)}\right)\ .
\eqno(6)$$

       The functional integral in the right-hand side of the Eq.(6)
     is taken in the space of 4-dimensional functions $\nu_{\mu}(\xi)$.  The
constant
     $ C_{\nu}$ is defined by the condition \\

$$ C_{\nu}\int{\delta^4{\nu}_{\mu}}exp\left(-i\int_{0}^{\tau}{d\xi}
      {[{\tilde g}^{\mu\nu}(x, \xi)]^{-1}}
 { {\nu}_{\mu}(\xi)}{{\nu}_{\nu}(\xi)} \right) = 1\ , $$

       from which it follows

$$
   C_{\nu} = \left[ \int{\delta^4{\nu}_{\mu}}
  exp\left( -i\int_{0}^{\tau}{d\xi}{[\tilde g^{\mu\nu}(x, \xi)]}\right)^{-1}
  {\nu}_{\mu}(\xi){\nu}_{\nu}(\xi)]\right]^{-1} =
    \left( det[\tilde g^{\mu\nu}(x, \xi)]\right)^{-1/2}\ .
$$

      After substituting (6) into (5), the operator $
exp\left(-2\int_{0}^{\tau}
    {{{\nu}^{\mu}(\xi)}{\partial_{\mu}(\xi)}{d\xi}}\right) $ can be
    "disentangled" and we can find  solution of the equation (4) in the form
     of the functional integral
$$ G(x, y|g^{\mu\nu}) =-i\int_{0}^{\infty}{d\tau}e^{-im^{2}{\tau}}\ \cdot\ $$
$$\ \cdot\  C_{\nu}\int{\delta^4{\nu}} exp\left(-im^2\int_{0}^{\tau}
[{\sqrt{-g(x_{\xi})}} -1]{d\xi}
-i\int_{0}^{\tau}{d\xi}[\tilde g^{\mu\nu}(x_{\xi})]^{-1}
{{\nu}_{\mu}(\xi)}{{\nu}_{\mu}(\xi)}\right)\ \cdot\ $$
$$\ \cdot\  {\delta}^{(4)}\left(x-y-2\int_{0}^{\tau}{\nu}(\eta){d\eta}\right),
\eqno(7)$$
where $$ x_{\xi} = x - 2\int_{\xi}^{\tau}{{\nu}(\eta)}{d\eta}\ . $$
      The Fourier transform of the Green function $ G(x, y/g^{\mu\nu}) $ has
the
following form
     $$ G(p, q|g^{\mu\nu}) =\int dx dy e^{ ipx-iqy } G(x, y/g^{\mu\nu}) =$$
$$ i\int_{0}^{\infty}e^{i(p^2 -m^2){\tau}}\int dy e^{i(p-q)y}\ \cdot\ $$
$$\ \cdot\
C_{\nu}\int{\delta}^4{\nu}exp\left(-im^{2}\int_{0}^{\tau}[\sqrt{-g(y_{\xi})}-1]d{\xi}
(-i)\int_{0}^{\tau}[\tilde g^{\mu\nu}(y_{\xi})]^{-1} [{\nu}(\xi)
+p]_{\mu}[{\nu}(\xi) + p]_{\nu} \right) \ ,\eqno(8)$$

$ y_{\xi} = y + \int_{0}^{\xi}[{\nu}(\eta) + p ]{d{\eta}}\ . $

      Eq.(8) is the closed expression for the scalar particle Green
     function in the external gravitational field.
     Let us now  consider the gravitational field in the linear
     approximation,  i. e  put  $ g_{\mu\nu} = \eta_{\mu\nu} + h_{\mu\nu} $
     where $ \eta_{\mu\nu} $ is the Minkowski metric tensor $ \eta_{\mu\nu}=
     diag ( 1,  -1,  -1,  -1 ) $.  Rewrite  Eq.(8) in the variables
     $ h_{\mu\nu} (x) $ after dropping out the terms with an exponent
     power higher than first $ h_{\mu\nu}(x) $\footnote[1]{Lagrangian (1)
     in the linear  approximation  to  $  h^{\mu\nu} $  has  the form
     $ L(x)= L_{0} + L_{int} $ ,   where

$$
L_{0} = \frac{1}{2}[\partial^{\mu}{\Psi}(x)\partial_{\mu}{\Psi}(x) -
m^{2}{\Psi}^{2}(x)]\ , $$
$$
L_{int}(x) = -\frac{1}{2}h^{\mu\nu}(x)T_{\mu\nu}, $$
$$
T_{\mu\nu}(x) = {\partial}_{\mu}{\Psi}(x){\partial}_{\nu}{\Psi}(x) -
\frac{1}{2}{\eta}_{\mu\nu}[ {\partial}^{\mu}{\Psi}(x){\partial}_{\mu}
{\Psi}(x) -m^{2}{\Psi}^{2}]\ ,$$
$T_{\mu\nu}(x)$-the  energy   momentum  tensor  of  the  scalar  field. }

$$ G(p, q|h^{\mu\nu}) =$$
$$ i\int_{0}^{\infty}{d\tau}e^{i(p^2-m^2)}
\int{dy}e^{i(p-q)y}
\int{\delta}^4{\nu}exp\left( -i\int_{0}^{\tau}{\nu}_{\mu}(\xi)
{{\nu}^{\mu}(\xi)}{d\xi} +i \int_{0}^{\tau} M(y_{\xi}){d\xi}\right)\ ,
\eqno(9)$$
where
$$ M(y_{\xi}) = \left( \frac{m^2}{2}h_{\nu}^{\nu}(y_{\xi}) +
[{\nu}(\xi) + p ]_{\mu}[{\nu}(\xi) + p]_{\nu}h^{\mu\nu}(y_{\xi}) \right)\ .
\eqno(10)$$

\section{Eikonal representation for the scattering amplitude of the
     scalar particle on the tensor potential}
      In this section we consider the scattering of the scalar particle in
     the external potential ${ h_{\mu\nu} }$. Using  expression (9) we find
    the scattering amplitude $  F(p, q|h_{\mu\nu}) $   / $ p $ and $ q$ are the
     particle momenta before and after the scattering respectively / by the
following
    formula:
$$ F(p, q|h^{\mu\nu}) = \lim_{p^2, q^2\to m^2}(p^2-m^2)(q^2-m^2)
G(p, q|h^{\mu\nu})\ . \eqno(11)$$

    Taking into account the identity  $ e^{a}-1 =
    a\int_{0}^{1}e^{\lambda a}d\lambda $,  we subtract from
    $ G(p, q|h^{\mu\nu}) $the free Green function
    $ G_{0}(p, q) =  { (2\pi)^{4}\delta^{4}(p-q)}/{p^2 -q^2} $
    which does not give  contribution to the scattering amplitude (11).
    As result,  we obtain
$$
F(p, q|h^{\mu\nu})=\lim_{p^2, q^2 \to m^2}(p^2-m^2)(q^2-m^2)
i\int_{0}^{\infty}e^{iTy}\int d{\tau}e^{i(p^2-m^2){\tau}}\ \cdot\ $$
$$
\ \cdot\ \int{\delta}^4{\nu}exp[ -i\int_{0}^{\tau}{\nu}_{\mu}(\xi)
{\nu}^{\mu}(\xi) d{\xi}] \int_{0}^{\tau} d{\tau} M(y_{\xi})\int_{0}^{1}
d{\lambda}e^{i{\lambda}\int_{0}^{\tau}d{\xi} M(y_{\xi})}\ , \eqno(12)$$

where $T = p-q$.

     Changing in Eq.(12) the variables

     $$ y= y' -2p{\eta} -2\int_{0}^{\eta}{\nu}(\eta'){d\eta'}\ ,$$
     $${\nu}(\xi)={\nu'}(\xi) -(p-q)\theta(\eta-\xi)\ , \eqno(13)$$

    $$\theta=\left\{\begin{array}{ll}
     1    & {\rm {\xi}}> 0 ,\\
     0    & {\rm {\xi}}< 0,
     \end{array}
\right. $$

   and using the relation $ [9] $
$$
\lim_{a, {\epsilon}\to 0}{ia\int_{0}^{\infty}{d\tau}
    e^{ia{\tau} - {\epsilon}{\tau}} f(\tau)} = f({\infty})\ , \eqno(14) $$

    we pass to the mass shell.  Finally the scattering amplitude has the
    form
$$F(p, q|h^{\mu\nu}) =$$
$$\int{dy}e^{iTy}\int{\delta^4}{\nu}exp(-i\int_{-\infty}^{\infty}
{\nu}_{\mu}(\xi){\nu}^{\mu}(\xi){d\xi})
M(y/0)\int_{0}^{1}d\lambda exp[i\lambda\int_{-\infty}^{\infty}{d\xi}
M(y/{\xi})]
,\eqno(15)$$
    where
$$
M(y|{\xi})=\frac{m^2}{2}h_{\nu}^{\mu}[y + 2p{\theta}(\xi){\xi} +
2q{\theta}(-\xi){\xi} + 2\int_{0}^{\xi}{\nu}(\eta){d\eta}] +   $$
$$
[ {\nu}(\xi) + p{\theta}(\xi) +q{\theta}(-\xi)]_{\mu}[{\nu}(\xi)
+ p{\theta}(\xi) + q{\theta}(-\xi) ]_{\nu}\ \cdot\ $$
$$\ \cdot\  h^{\mu\nu}[y +
2p{\theta}(\xi) +2q{\theta}(-\xi){\xi} + 2\int_{0}^{\xi}{\nu}(\eta)
{d\eta}] \ .\eqno(16)$$

    For the calculation of the functional integrals we use the straight line
    path approximation $ [3, 4, 10]  $, i. e. we  assume that in the high
energy
    particle scattering  on the smooth potential $ h^{\mu\nu} $,  one can
    be neglect the dependence on the functional variables
    ${\nu}_{\mu}(\eta) $.  In other words,  it is considered that the main
    contribution to the functional integral (15) comes from a
    trajectory particle moving freely from the momentum ${\vec p} $
    with  ${\xi}> {0} $  to the momentum ${\vec q} $
    with  ${\xi}< {0} $ and passing via the point $ y $ with
    ${\xi} = 0 $.

   To simplify,  we consider the case when the potential does
      not depend on the time:$ h^{\mu\nu}(y) = h^{\mu\nu}(\vec r, t)=h^{\mu\nu}
$.
    Therefore,  for the scattering amplitude we obtain the
    following closed expression\footnote[2]{ the amplitude $ f(p, q) $ is
    normalized by the relations $$ {\sigma}=\frac{4\pi}{\mid\vec
p\mid}Imf(p, p), \frac{d{\sigma}}{d{\Omega}}= |f(p, q)|^{2}. $$ }

$$F(p, q) =2(2\pi)^2{\delta}(p_0-q_0)f(p, q),$$

$$ f(p, q)= \frac{1}{4\pi}\int{d\vec r}e^{iT\vec r} M'({\vec r}/0)
   \int_{0}^{1}d\lambda exp[i{\lambda}{\chi}(\vec r)] ,\eqno(17)$$
 where
 $$ M'({\vec r}|0)=\left( \frac{m^2}{2}h_{\nu}^{\nu}(\vec r) +
\frac{1}{4} [p+q]_{\mu}[p+q]_{\nu}h^{\mu\nu}(\vec r)\right)\ , $$
$$
{\chi}_0(\vec r)=
\frac{1}{2\mid\vec p\mid}\int_{-\infty}^{\infty}{ds}[(\frac{m^2}{2}
h_{\nu}^{\nu}({\vec r} +{\hat p}{\theta}(s)s +{\hat q}{\theta}(-s)s )]+ $$
$$\frac{1}{2\mid\vec p\mid}
\int_{-\infty}^{\infty}{ds}\left([p{\theta}(s)+q{\theta}(-s)]_{\mu}
[p{\theta}(s) + q{\theta}(-s)]_{\nu}
h^{\mu\nu}({\vec r} +p{\theta}(s)s+\hat{q}{\theta}(-s)s) \right)\ , \eqno(18)
$$

${\hat p} = \frac{\mid\vec p\mid}{| p|} $.
     In the case of  weak gravitational field, the tensor $ h^{\mu\nu} $
     has the form $ [11] $
$$
h^{00}(\vec r) =2{\phi(\vec r)}\ , $$
$$
h^{\alpha\beta}(\vec r) = 2{\delta}_{\alpha\beta}{\phi}(\vec r)\ , \eqno(19) $$
$$
h^{0\alpha}(\vec r) =h^{\alpha 0}(\vec r) = 0\ , $$
$$
({\alpha}, {\beta} = 1, 2, 3 ).  $$

    Since we consider the high energy scattering,   the first term in Eq.(18)
 for $\chi_0 (\vec r)$ can be neglected in comparison with the second one.  As
    a result,  we have
$$ f(p, q)=\frac{1}{2\pi}\int{d\vec r}e^{iT\vec
r}(p_{0}^2 + {\vec p}^2) {\phi(\vec
r)}\int_{0}^{1}{d\lambda}exp(i{\lambda}{\chi_{0}(\vec r)})\ ,\eqno(20)$$

    where
$$
{\chi}_0(\vec r) = \frac{1}{\mid\vec p\mid}\int_{-\infty}^{\infty}{ds}
(p^2+{\vec p}^2)\phi({\vec r} + {\hat p}s)\ .\eqno(21) $$

    We direct the axis $z$ along the momentum ${\vec p}$,  but  we place the
$x$
    axis  into the plane defined by the vectors ${\vec p}$ and ${\vec q}$.
   In this coordinate system,  for the phase (21) in the presence of small
angles ${\theta}\ll 1$  we obtain the following expression:
$$ {\chi}_{0}(p_{0}, {\vec r})
= \frac{2p_{0}^{2}}{{\mid\vec p_z\mid}}
\left(\int_{0}^{\infty}{ds}{\phi}(x, y, z+s) + \int_{-\infty}^{0}ds
{\phi}(x+s\sin{\theta},  y,  z+s\cos{\theta} )\right){\cong}
$$
$$
=\frac{2p_{0}^{2}}{{\mid\vec p_z\mid}}\int_{-\infty}^{\infty}{ds}
{\phi}(x, y, z+s)=\frac{2p_{0}^{2}}{{\mid \vec p_z\mid}}\int_{-\infty}^{\infty}
{dz}{\phi}(\vec r) \ .\eqno(22)$$

    At small angles  the momentum transfer is nearly
     perpendicular to the $ z $-axis therefore,  in Eq.(20) it
     can  put $ exp ({i{\vec T}{\vec r}}) = exp {i(T_{x}x + T_{y}y )} $.
    Integrating over $ {dz} $ and ${ d{\lambda}} $ with taking
    account of (22), we find the Glauber-type representation for the
    scattering amplitude [13]
$$
f(p, q) = -\frac{i{\mid\vec p_z\mid}}{2\pi}\int {d^2}{\vec b}
e^{i{\vec b}{\vec T_{\bot}}}\left({\exp[i{\chi}_0(p_0, {\vec b})]- 1}\right)\ ,
\eqno(23)$$

    where $ {\vec b} = ( x, y, 0) $,  and the eikonal phase
    $ {\chi}_{0}(p_0, b) $ is determined by  Eq.(22).
      Let us consider the Newton potential as limit of the Yukawa potential
    when $ {\mu} \to 0 $,   ${\phi}(\vec r)
=\left({(kMe^{-{\mu}r)}/{r}}\right)_
    {{\mu} = 0} =(-{kM})/{r} $, where $ k $ is the gravitational constant,
    $ k = 6. 10^{-39}m_{p}^{2}$ and    $ M $ the mass creating the potential.
In
this
    case an eikonal phase  (22) get the following form
$$
{\chi}_{0}(p_{0}, {\vec b}) = -\frac{2kp_{0}^{2}M}{{\mid\vec p_z\mid}}
 \int_{-\infty}^{\infty}{dz}\frac{e^{-{\mu}\sqrt{b^2+z^2}}}
  {\sqrt{b^2 + z^2}} =-{\beta}K_{0}({\mu}|b)\ , \eqno(24)$$

    where $ \beta=\frac{kM{p_0}^2}{2\pi{\mid\vec p_z\mid}} $,
$ K_0(\mu|b)=\frac{1}{2\pi}\int{d^2\vec{ k_\bot}}
\frac{e^{i {\vec k_\bot}\vec b_\bot}}
{{\vec k_\bot}^2 +\mu^2 }$-is the Kelvin function of the zeroth order.
For the scattering amplitude,  we have the following expression:
$$
f(p, q) = -\frac{i{\mid \vec p_z\mid}}{2\pi}\int{d^2{\vec b}}
e^{i{\vec b}{\vec T}}\left(exp[-i\frac{\beta}{2\pi}\int{d^2k_{\bot}}
\frac{e^{i{\vec k}_{\bot}{\vec b}_{\bot}}}{{\vec k}_{\bot}^2 + {\mu}^2}]
-1\right)\ .\eqno(25) $$

       Calculate the integral (25), preserving only the terms which
    does not disappear when $ {\mu}\to  0 $

$$ f({\vec p}{\vec q}) = - \frac{kMp_{0}^{2}}{{\pi}t}
   \frac{{\Gamma}(1+i{\beta})}{{\Gamma}(1-i{\beta})}
   exp[ -i{\beta}ln(\frac{\sqrt{t}}{{\mu}c})]\ , \eqno(26)$$
    where ${t=-{\vec T}_{\bot}^2}$; ${C}$ is the Euler constant.   The phase
    diverging  when ${ \mu}\to 0 $ in (26) caused,  it is well known,  by
    the long-range of action of the Newton potential [14].  From comparing (26)
with the result of the work [10], devoted to the consideration
    of scattering of the scalar particle by the vector potential, one
    may conclude that
    $ {{\sigma}_{grav. }}/{{\sigma}_{vec. }}\sim ({k^2M^2}/{e^2})p_0^2 $.
    The poles of the amplitude determined by  Eq.(26)  give the
    discrete energy levels of particles in the Newton potential
$$
E_{n} =-\frac{{k^2m^2}M(m+M)}{8{\pi}^2}\frac{1}{n^2}\ , \eqno(27)$$
$$ n= 1, 2, 3,...  $$.

    If in the Eq.(27)  we put $ m=M \sim m_{nuclon} $,
    we obtain the energy of the ground state  equal to

 $$ E_{1} = 9, 4. {10}^{-70} eV \ . $$

\section{Acknowledgements}
     I am grateful to Profs. B.M. Barbashov, G.V. Efimov, A.V. Efremov, M.K.
Volkov, V.V. Nesterenko, V.N. Pervushin for useful discussions. I am also
indebted to Prof. P. Chiappetta for financial support during my stay at the
Centre de Physique Th\'eorique in Marseille and warm hospitality.

\end{document}